\pdfoutput=1
\RequirePackage{ifpdf}
\ifpdf 
\documentclass[pdftex]{sigma}
\else
\documentclass{sigma}
\fi

\begin{document}


\newcommand{\arXivNumber}{1601.01181}

\renewcommand{\PaperNumber}{027}

\FirstPageHeading

\ShortArticleName{A Simple Proof of Sklyanin's Formula}

\ArticleName{A Simple Proof of Sklyanin's Formula\\ for Canonical Spectral Coordinates\\ of the Rational Calogero--Moser System}

\Author{Tam\'as F.~G\"ORBE}

\AuthorNameForHeading{T.F.~G\"orbe}

\Address{Department of Theoretical Physics, University of Szeged,\\ Tisza Lajos krt 84-86, H-6720 Szeged, Hungary}
\Email{\href{mailto:tfgorbe@physx.u-szeged.hu}{tfgorbe@physx.u-szeged.hu}}
\URLaddress{\url{http://www.staff.u-szeged.hu/~tfgorbe/}}

\ArticleDates{Received January 19, 2016, in f\/inal form March 08, 2016; Published online March 11, 2016}

\Abstract{We use Hamiltonian reduction to simplify Falqui and Mencattini's recent proof
of Sklyanin's expression providing spectral Darboux coordinates of the rational
Calogero--Moser system. This viewpoint enables us to verify a conjecture of
Falqui and Mencattini, and to obtain Sklyanin's formula as a corollary.}


\Classification{14H70; 37J15; 53D20}

\section{Introduction}\label{sec:1}

Integrable many-body systems in one spatial dimension form an important class of
exactly solvable Hamiltonian systems with their diverse mathematical structure and
widespread applicability in physics \cite{Ca01,Pe90,Su04}. Among these many-body
systems, one of the most widely known is the rational Calogero--Moser model of equally
massive interacting particles moving along a~line with a pair potential inversely
proportional to the square of the distance. The model was introduced and solved at
the quantum level by Calogero~\cite{Ca71}. The complete integrability of its classical
version was established by Moser~\cite{Mo75}, who employed the Lax formalism to identify
a~complete set of commuting integrals as coef\/f\/icients of the characteristic polynomial
of a~certain Hermitian matrix function, called the Lax matrix.

These developments might prompt one to consider the Poisson commuting eigenvalues
of the Lax matrix and be interested in searching for an expression of conjugate variables.
Such an expression was indeed formulated by Sklyanin~\cite{Sk09} in his work on
bispectrality, and worked out in detail for the open Toda chain \cite{Sk13}.
Sklyanin's formula for the rational Calogero--Moser model was recently conf\/irmed
within the framework of bi-Hamiltonian geometry by Falqui and Mencattini~\cite{FM15}
in a somewhat circuitous way, although a short-cut was pointed out in the form of a~conjecture.
The purpose of this paper is to prove this conjecture and of\/fer
an alternative simple proof of Sklyanin's formula using results of Hamiltonian reduction.

Section~\ref{sec:2} is a recap of complete integrability and action-angle duality
for the rational Calo\-ge\-ro--Mo\-ser system in the context of Hamiltonian reduction.
In Section~\ref{sec:3} we put these ideas into practice when we identify the canonical
variables of~\cite{FM15} in terms of the reduction picture, and prove
the relation conjectured in that paper. We attain Sklyanin's formula as a corollary.
Section~\ref{sec:4} contains our concluding remarks on possible generalizations.

\section{The rational Calogero--Moser system via reduction}
\label{sec:2}

We begin by describing the rational Calogero--Moser system and recalling how it originates
from Hamiltonian reduction~\cite{KKS78}. The content of this section is standard and only
included for the sake of self-consistency.

For $n$ particles, let the $n$-tuples $q=(q_1,\dots,q_n)$ and $p=(p_1,\dots,p_n)$ collect
their coordinates and momenta, respectively. Then the Hamiltonian of the model reads
\begin{gather}
H(q,p)=\frac{1}{2}\sum_{j=1}^np_j^2
+g^2\sum_{\substack{j,k=1\\(j<k)}}^n\frac{1}{(q_j-q_k)^2},
\label{1}
\end{gather}
where $g$ is a real coupling constant tuning the strength of particle interaction.
The pair potential is singular at $q_j=q_k$ $(j\neq k)$, hence any initial ordering
of the particles remains unchanged during time-evolution. The conf\/iguration space is
chosen to be the domain $\mathcal{C}=\{q\in\mathbb{R}^n\,|\, q_1>\dots>q_n\}$, and the phase space is
its cotangent bundle
\begin{gather}
T^\ast\mathcal{C}=\big\{(q,p)\,|\, q\in\mathcal{C},\, p\in\mathbb{R}^n\big\},
\label{2}
\end{gather}
endowed with the standard symplectic form
\begin{gather}
\omega=\sum_{j=1}^n dq_j\wedge dp_j.
\label{3}
\end{gather}
The Hamiltonian system $(T^\ast\mathcal{C},\omega,H)$, called the rational Calogero--Moser
system, can be obtained as an appropriate Marsden--Weinstein reduction of the free
particle moving in the space of $n\times n$ Hermitian matrices as follows.

Consider the manifold of pairs of $n\times n$ Hermitian matrices
\begin{gather}
M=\big\{(X,P)\,|\, X,P\in\mathfrak{gl}(n,\mathbb{C}),\, X^\dag=X,\, P^\dag=P\big\},
\label{4}
\end{gather}
equipped with the symplectic form
\begin{gather}
\Omega=\operatorname{tr}(dX\wedge dP).
\label{5}
\end{gather}
The Hamiltonian of the analogue of a free particle reads
\begin{gather*}
\mathcal{H}(X,P)=\frac{1}{2}\operatorname{tr}\big(P^2\big).
\end{gather*}
The equations of motion can be solved explicitly for this Hamiltonian system
$(M,\Omega,\mathcal{H})$, and the general solution is given by $X(t)=tP_0+X_0$,
$P(t)=P_0$. Moreover, the functions $\mathcal{H}_k(X,P)=\frac{1}{k}\operatorname{tr}\big(P^k\big)$, $k=1,\dots,n$
form an independent set of commuting f\/irst integrals.

The group of $n\times n$ unitary matrices $U(n)$ acts on $M$~\eqref{4} by conjugation
\begin{gather*}
(X,P)\to \big(UXU^\dag,UPU^\dag\big),\qquad U\in U(n),
\end{gather*}
leaves both the symplectic form $\Omega$ \eqref{5} and the Hamiltonians $\mathcal{H}_k$
invariant, and the matrix commutator $(X,P)\to[X,P]$ is a momentum map for this
$U(n)$-action. Consider the Hamiltonian reduction performed by factorizing the
momentum constraint surface
\begin{gather*}
[X,P]=\mathrm{i} g\big(vv^\dag-\mathbf{1}_n\big)=:\mu, \qquad v=(1\dots 1)^\dag\in\mathbb{R}^n,\qquad g\in\mathbb{R},
\end{gather*}
with the stabilizer subgroup $G_\mu\subset U(n)$ of~$\mu$, e.g., by diagonalization of
the~$X$ component. This yields the gauge slice $S=\{(Q(q,p),L(q,p))\,|\, q\in\mathcal{C},\, p\in\mathbb{R}^n\}$,
where
\begin{gather}
Q_{jk}=(UXU^\dag)_{jk}=q_j\delta_{jk},\nonumber\\
L_{jk}=(UPU^\dag)_{jk}=p_j\delta_{jk}+\mathrm{i} g\frac{1-\delta_{jk}}{q_j-q_k},\qquad
j,k=1,\dots,n.
\label{9}
\end{gather}
This $S$ is symplectomorphic to the reduced phase space and to $T^\ast\mathcal{C}$
\eqref{2} since it inherits the reduced symplectic form $\omega$~\eqref{3}.
The unreduced Hamiltonians project to a commuting set of independent integrals
$H_k=\frac{1}{k}\operatorname{tr}\big(L^k\big)$, $k=1,\dots,n$, such that $H_2=H$~\eqref{1} and what's more,
the completeness of Hamiltonian f\/lows follows automatically from the reduction.
Therefore the rational Calogero--Moser system is completely integrable.

The similar role of matrices $X$ and $P$ in the derivation above can be exploited to
construct action-angle variables for the rational Calogero--Moser system. This is done by
switching to the gauge, where the $P$ component is diagonalized by some matrix
$\tilde U\in G_\mu$, and it boils down to the gauge slice
$\tilde S=\big\{\big(\tilde Q(\phi,\lambda),\tilde L(\phi,\lambda)\big)\,|\,\phi\in\mathbb{R}^n,\, \lambda\in\mathcal{C}\big\}$, where
\begin{gather}
\tilde Q_{jk}=\big(\tilde UX\tilde U^\dag\big)_{jk}
=\phi_j\delta_{jk}-\mathrm{i} g\frac{1-\delta_{jk}}{\lambda_j-\lambda_k},\nonumber\\
\tilde L_{jk}=\big(\tilde UP\tilde U^\dag\big)_{jk}=\lambda_j\delta_{jk},\qquad
j,k=1,\dots,n.
\label{10}
\end{gather}
By construction, $\tilde S$ with the symplectic form
$\tilde\omega=\sum\limits_{j=1}^nd\phi_j\wedge d\lambda_j$
is also symplectomorphic to the reduced phase space, thus a canonical transformation
$(q,p)\to(\phi,\lambda)$ is obtained, where the reduced Hamiltonians depend only on
$\lambda$, viz.\ $H_k=\frac{1}{k}\big(\lambda_1^k+\dots+\lambda_n^k\big)$, $k=1,\dots,n$.

\section{Sklyanin's formula}
\label{sec:3}

Now, we turn to the question of variables conjugate to the Poisson commuting eigenvalues
$\lambda_1,\dots,\lambda_n$ of $L$~\eqref{9}, i.e., such functions
$\theta_1,\dots,\theta_n$ in involution that
\begin{gather*}
\{\theta_j,\lambda_k\}=\delta_{jk},\qquad j,k=1,\dots,n.
\end{gather*}
At the end of Section~\ref{sec:2} we saw that the variables $\phi_1,\dots,\phi_n$ are
such functions. These action-angle variables~$\lambda$,~$\phi$ were already obtained by
Moser~\cite{Mo75} using scattering theory, and also appear in Ruijsenaars' proof of
the self-duality of the rational Calogero--Moser system~\cite{Ru88}.

Let us def\/ine the following functions over the phase space $T^\ast\mathcal{C}$
\eqref{2} with dependence on an additional variable~$z$:
\begin{gather}
A(z)=\det(z\mathbf{1}_n-L),\qquad
C(z)=\operatorname{tr}\big(Q\operatorname{adj}(z\mathbf{1}_n-L)vv^\dag\big),\nonumber\\
D(z)=\operatorname{tr}\big(Q\operatorname{adj}(z\mathbf{1}_n-L)\big),
\label{12}
\end{gather}
where $Q$ and $L$ are given by \eqref{9}, $v=(1\dots1)^\dag\in\mathbb{R}^n$ and $\operatorname{adj}$ denotes the
adjugate matrix, i.e., the transpose of the cofactor matrix.
Sklyanin's formula~\cite{Sk09} for $\theta_1,\dots,\theta_n$ then reads
\begin{gather}
\theta_k=\frac{C(\lambda_k)}{A'(\lambda_k)},\qquad k=1,\dots,n.
\label{13}
\end{gather}
In \cite{FM15} Falqui and Mencattini have shown that
\begin{gather}
\mu_k=\frac{D(\lambda_k)}{A'(\lambda_k)},\qquad k=1,\dots,n
\label{14}
\end{gather}
are conjugate variables to $\lambda_1,\dots,\lambda_n$, and
\begin{gather}
\theta_k=\mu_k+f_k(\lambda_1,\dots,\lambda_n),\qquad k=1,\dots,n,
\label{15}
\end{gather}
with such $\lambda$-dependent functions $f_1,\dots,f_n$ that
\begin{gather}
\frac{\partial f_j}{\partial\lambda_k}=\frac{\partial f_k}{\partial\lambda_j},\qquad
j,k=1,\dots,n
\label{16}
\end{gather}
thus $\theta_1,\dots,\theta_n$ given by
Sklyanin's formula~\eqref{13} are conjugate to $\lambda_1,\dots,\lambda_n$.
This was done in a~roundabout way, although the explicit form of
relation~\eqref{15} was conjectured.

Here we take a dif\/ferent route by making use of the reduction viewpoint of Section~\ref{sec:2}. From this perspective, the problem becomes transparent and can be solved
ef\/fortlessly. First, we show that $\mu_1,\dots,\mu_n$~\eqref{14} are nothing else
than the angle variables $\phi_1,\dots,\phi_n$.

\begin{lemma*}
The variables $\mu_1,\dots,\mu_n$ defined in~\eqref{14} are the angle variables
$\phi_1,\dots,\phi_n$ of the rational Calogero--Moser system.
\end{lemma*}

\begin{proof}
Notice that, by def\/inition, $\mu_1,\dots,\mu_n$ are gauge invariant, thus by working in
the gauge, where the $P$ component is diagonal, that is with the matrices~$\tilde Q$,~$\tilde L$~\eqref{10}, we get
\begin{gather}
\frac{D(z)}{A'(z)}
=\frac{\sum\limits_{j=1}^n\phi_j\prod\limits_{\substack{\ell=1\\ (\ell\neq j)}}^n(z-\lambda_\ell)}
{\sum\limits_{j=1}^n\prod\limits_{\substack{\ell=1\\ (\ell\neq j)}}^n(z-\lambda_\ell)}.
\label{17}
\end{gather}
Substituting $z=\lambda_k$ into~\eqref{17} yields $\mu_k=\phi_k$, for each $k=1,\dots,n$.
\end{proof}

Next, we prove the relation of functions $A$, $C$, $D$ \eqref{12}, that was
conjectured in \cite{FM15}.

\begin{theorem*}
For any $n\in\mathbb{N}$, $(q,p)\in T^\ast\mathcal{C}$ \eqref{2}, and $z\in\mathbb{C}$ we have
\begin{gather*}
C(z)=D(z)+\frac{\mathrm{i} g}{2}A''(z).
\end{gather*}
\end{theorem*}

\begin{proof}
Pick any point $(q,p)$ in the phase space $T^\ast\mathcal{C}$ and consider the
corresponding point $(\lambda,\phi)$ in the space of action-angle variables.
Since $A(z)=(z-\lambda_1)\cdots(z-\lambda_n)$ we have
\begin{gather*}
\frac{\mathrm{i} g}{2}A''(z)
=\mathrm{i} g\sum_{\substack{j,k=1\\ (j<k)}}^n
\prod_{\substack{\ell=1\\(\ell\neq j,k)}}^n(z-\lambda_\ell).
\end{gather*}
The dif\/ference of functions $C$ and $D$ \eqref{12} reads
\begin{gather}
C(z)-D(z)=\operatorname{tr}\big(Q\operatorname{adj}(z\mathbf{1}_n-L)\big(vv^\dag-\mathbf{1}_n\big)\big).
\label{20}
\end{gather}
Due to gauge invariance, we are allowed to work with~$\tilde Q$,~$\tilde L$~\eqref{10}
instead of~$Q$,~$L$~\eqref{9}. Therefore~\eqref{20} can be written as the sum of all
of\/f-diagonal components of $\tilde Q\operatorname{adj}(z\mathbf{1}_n-\tilde L)$, that is
\begin{gather*}
C(z)-D(z)=\mathrm{i} g\sum_{\substack{j,k=1\\ (j\neq k)}}^n\frac{-1}{\lambda_j-\lambda_k}
\prod_{\substack{\ell=1\\(\ell\neq k)}}^n(z-\lambda_\ell)=
\mathrm{i} g\sum_{\substack{j,k=1\\ (j<k)}}^n
\prod_{\substack{\ell=1\\ (\ell\neq j,k)}}^n(z-\lambda_\ell).
\end{gather*}
This concludes the proof.
\end{proof}

Our theorem conf\/irms that indeed relation \eqref{15} is valid with
\begin{gather*}
f_k(\lambda_1,\dots,\lambda_n)
=\frac{\mathrm{i} g}{2}\frac{A''(\lambda_k)}{A'(\lambda_k)}
=\mathrm{i} g\sum_{\substack{\ell=1\\(\ell\neq k)}}^n\frac{1}{\lambda_k-\lambda_\ell}
,\qquad
k=1,\dots,n,
\end{gather*}
for which \eqref{16} clearly holds. An immediate consequence, as we indicated before,
is that $\theta_1,\dots,\theta_n$ \eqref{13} are conjugate variables to
$\lambda_1,\dots,\lambda_n$, thus Sklyanin's formula is verif\/ied.

\begin{corollary*}[Sklyanin's formula]
The variables $\theta_1,\dots,\theta_n$ defined by
\begin{gather*}
\theta_k=\frac{C(\lambda_k)}{A'(\lambda_k)},\qquad k=1,\dots,n
\end{gather*}
are conjugate to the eigenvalues $\lambda_1,\dots,\lambda_n$ of the Lax
matrix~$L$.
\end{corollary*}

\section{Discussion}\label{sec:4}

There seem to be several ways for generalization. For example, one might consider
rational Calogero--Moser models associated to root systems other than type~$A_{n-1}$.
The hyperbolic Calo\-gero--Moser systems as well as, the `relativistic' Calogero--Moser
systems, also known as Ruijsenaars--Schneider systems, are also of considerable interest.

\subsection*{Acknowledgements}

Thanks are due to L\'aszl\'o Feh\'er for drawing our attention to the duality
perspective. This work was supported in part by the Hungarian Scientif\/ic Research
Fund (OTKA) under the grant K-111697. The work was also partially supported by COST
(European Cooperation in Science and Technology) in COST Action MP1405 QSPACE.

\pdfbookmark[1]{References}{ref}
\LastPageEnding


\begin{thebibliography}{99}
\footnotesize\itemsep=0pt

\bibitem{Ca71}
Calogero F., Solution of the one-dimensional {$N$}-body problems with quadratic
 and/or inversely quadratic pair potentials, \href{http://dx.doi.org/10.1063/1.1665604}{\textit{J.~Math. Phys.}}
 \textbf{12} (1971), 419--436, {E}rratum,
 \href{http://dx.doi.org/10.1063/1.531804}{\textit{J.~Math. Phys.}}
 \textbf{37} (1996), 3646.

\bibitem{Ca01}
Calogero F., Classical many-body problems amenable to exact treatments,
 \href{http://dx.doi.org/10.1007/3-540-44730-X}{\textit{Lecture Notes in Physics. New Series~m: Monographs}}, Vol.~66,
 Springer-Verlag, Berlin, 2001.

\bibitem{FM15}
Falqui G., Mencattini I., Bi-Hamiltonian geometry and canonical spectral
 coordinates for the rational {C}alogero--{M}oser system, \href{http://arxiv.org/abs/1511.06339}{arXiv:1511.06339}.

\bibitem{KKS78}
Kazhdan D., Kostant B., Sternberg S., Hamiltonian group actions and dynamical
 systems of {C}alogero type, \href{http://dx.doi.org/10.1002/cpa.3160310405}{\textit{Comm. Pure Appl. Math.}} \textbf{31}
 (1978), 481--507.

\bibitem{Mo75}
Moser J., Three integrable {H}amiltonian systems connected with isospectral
 deformations, \href{http://dx.doi.org/10.1016/0001-8708(75)90151-6}{\textit{Adv. Math.}} \textbf{16} (1975), 197--220.

\bibitem{Pe90}
Perelomov A.M., Integrable systems of classical mechanics and {L}ie algebras.
 {V}ol.~{I}, \href{http://dx.doi.org/10.1007/978-3-0348-9257-5}{Birkh\"auser Verlag}, Basel, 1990.

\bibitem{Ru88}
Ruijsenaars S.N.M., Action-angle maps and scattering theory for some
 f\/inite-dimensional integrable systems. {I}.~{T}he pure soliton case,
 \href{http://dx.doi.org/10.1007/BF01238855}{\textit{Comm. Math. Phys.}} \textbf{115} (1988), 127--165.

\bibitem{Sk09}
Sklyanin E., Bispectrality and separation of variables in multiparticle
 hypergeometric systems, {T}alk given at the Workshop `Quantum Integrable
 Discrete Systems', Cambridge, England, March 23--27, 2009.

\bibitem{Sk13}
Sklyanin E., Bispectrality for the quantum open {T}oda chain,
 \href{http://dx.doi.org/10.1088/1751-8113/46/38/382001}{\textit{J.~Phys.~A: Math. Theor.}} \textbf{46} (2013), 382001, 8~pages,
 \href{http://arxiv.org/abs/1306.0454}{arXiv:1306.0454}.

\bibitem{Su04}
Sutherland B., Beautiful models: 70 years of exactly solved quantum many-body
 problems, \href{http://dx.doi.org/10.1142/5552}{World Sci. Publ. Co., Inc.}, River Edge, NJ, 2004.

\end{thebibliography}
\end{document}